\documentstyle[twocolumn,prl,aps,epsfig]{revtex}

\title{Quantum Antidots: Coulomb Blockade or no Coulomb Blockade?}

\begin{document}
\newcommand{\showfigures}{no}     
\maketitle
In a recent Letter\cite{Kataoka} Kataoka {\em et al.} critique our statement "... there is no Coulomb Blockade for resonant tunneling through an antidot since there is no isolated region that is being charged."\cite{SciCharge,Maasilta} They furnish their "proof" of Coulomb Blockade (CB) in quantum antidots (QAD) in the integer quantum Hall regime by means of an intricate experiment: they sense the variation of the fringe electric field when the electron occupation of a QAD changes as a function of applied magnetic field.

While using or not using the terminology "Coulomb Blockade" is a matter of semantics to a large extent, the common usage of the term \cite{GiaeverAverin} refers to charging of an isolated metallic island where: (i) the electron energy spectrum is continuous without the CB, and (ii) the charging energy can be conveniently expressed as $U_C = e^2/2C$ with $C$ being the total capacitance of the island to the outside world. Here "isolated" is important because an elecrical insulator defines conduction electron vacuum, and thus allows us to (i) define the metallic island, and (ii) to fix the number of electrons therein. The notion of CB is most useful when $C$ is constant (so long as just a few electrons are added or subtracted), and can be easily found from the geometry. The notion of CB is less useful even for single electron tunneling in quantum dots, where $U_C$ acquires additional large contributions, such as the size quantization energy and the intradot Coulomb interaction energy, both being of the same order as the geometric charging energy. \cite{SciDot} Here the point is that the physics changes from essentially single particle in metallic islands to the many body in true quantum dots; the simple single particle models give numerically inaccurate energies, and neglect new class of many body effects, such as the spin singlet - triplet transitions of a two electron state in a quantum dot. \cite{QDspin}

\begin{figure}[h]
\begin{center}
\epsfig{file=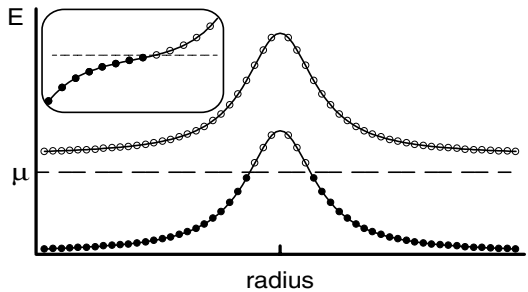, width=7cm}
\caption{A schematic energy profile of a quantum antidot in quantum Hall regime. The electron states of a Landau level (circles) are quantized by the Aharonov-Bohm condition that the state number $m$, starting from the center, contains $m \phi_0$ magnetic flux. At a low temperature $k_BT \ll \Delta E$ only states below chemical potential $\mu$ are occupied. Inset shows the self consistent energy profile near $\mu$. The Figure illustrates that there is no boundary separating the antidot bound states, experiencing CB, from other electron states. }
\label{Fig. 1}
\end{center}
\end{figure}

The notion of CB is even less useful for QADs, and is even conceptually ill defined as no vacuum separates the electrons bound on the QAD from all the rest of electrons in the system. That is, it is not clear which electrons are to be considered as bound on the QAD and which are not (Fig. 1). The only natural criterion is to consider the energy spectrum as discrete (electrons bound on the QAD) for $\Delta E$ greater than temperature and any external excitation, and continuous otherwise. Thus we are forced to conclude that the "size" of the QAD depends on temperature and on voltage used to measure conductance. In addition, if one were to estimate the geometric CB charging energy for a QAD, \cite{Numbers} one obtains $\sim$12 meV for QAD of Ref. \cite{Kataoka} and $\sim$4 meV for QAD of Ref. \cite{Maasilta}; these values are some 200 times greater than the experimental level spacing $\Delta E$ obtained from thermal activation in both works. 

Further, all experimental results of Ref. \cite{Kataoka} simply tell us that the electron energy spectrum on the QAD is discrete, and that electrons respond to electric field. However, a discrete energy spectrum by itself is {\em not} commonly taken to imply Coulomb Blockade. For example, atomic and molecular energy spectra are discrete, yet standard texts do not attribute this to CB.

\vspace{12pt}
\noindent V. J. Goldman

Department of Physics

SUNY at Stony Brook

Stony Brook, NY 11794-3800


\end{document}